# Building Resilient SMEs: Harnessing Large Language Models for Cyber Security in Australia

**Ben Kereopa-Yorke, Author**

*The escalating digitalisation of our lives and enterprises has led to a parallel growth in the complexity and frequency of cyber-attacks. Small and medium-sized enterprises (SMEs), particularly in Australia, are experiencing increased vulnerability to cyber threats, posing a significant challenge to the nation's cyber security landscape. Embracing transformative technologies such as Artificial Intelligence (AI), Machine Learning (ML) and Large Language Models (LLMs) can potentially strengthen cyber security policies for Australian SMEs. However, their practical application, advantages, and limitations remain underexplored, with prior research mainly focusing on large corporations. This study aims to address this gap by providing a comprehensive understanding of the potential role of LLMs in enhancing cyber security policies for Australian SMEs. Employing a mixed-methods study design, this research includes a literature review, qualitative analysis of SME case studies, and a quantitative assessment of LLM performance metrics in cyber security applications. The findings highlight the promising potential of LLMs across various performance criteria, including relevance, accuracy, and applicability, though gaps remain in areas such as completeness and clarity. The study underlines the importance of integrating human expertise with LLM technology and refining model development to address these limitations. By proposing a robust conceptual framework guiding the effective adoption of LLMs, this research aims to contribute to a safer and more resilient cyber environment for Australian SMEs, enabling sustainable growth and competitiveness in the digital era.*

## Introduction

In the face of escalating digitalisation, the incidence and complexity of cyber-attacks have concurrently soared. Small and medium-sized enterprises (SMEs), pivotal to Australia's economy, frequently fall victim to such cyber threats, posing a profound challenge to the cybersecurity landscape. Traditional cyber security methods often grapple with the rapidly evolving threat landscape and the pace of technological advancement, necessitating a shift towards transformative technologies such as Artificial Intelligence (AI), Machine Learning (ML) and Large Language Models (LLMs).

AI's capacity to process immense amounts of data in real-time, learn from it, and anticipate threats can play a game-changing role in proactively confronting cyber threats. LLMs, a particular subset of AI, have the potential to revolutionise the interpretation of unstructured data, which is frequently sidelined in conventional cyber security approaches. Nevertheless, the integration of AI and LLMs into cyber security also introduces a host of challenges, including novel attack vectors, concerns around data privacy, transparency, explainability, accountability, and specific hurdles related to cost, technical expertise, and regulatory issues within SMEs.

Despite a burgeoning body of literature on AI, ML and LLMs, their practical applications, benefits, and limitations within the context of cyber security policies for SMEs remain underexplored. Prior research has predominantly focused on large corporations, while the unique challenges and needs of SMEs have received less attention. Moreover, there is a shortage of studies addressing the potential application of LLMs in cyber security for Australian SMEs, a gap this study seeks to address.

This research, titled "Building Resilient SMEs: Harnessing Large Language Models for Cyber Security in Australia", navigates these intricacies. It aspires to provide a comprehensive understanding of the potential role of LLMs in enhancing cyber security policies within Australian SMEs. The investigation is guided by eight core research questions, focusing on the challenges in formulating effective cybersecurity policies, the practical implications of LLMs, their potential advantages and limitations, and future research directions.

The primary goal of the study is to propose a robust conceptual framework that guides the effective adoption of LLMs, thereby contributing to a safer and more resilient cyber environment for Australian SMEs. By offering a comprehensive exploration of the prospects of LLMs in cyber security, this study is intended to ensure the sustained growth and competitiveness of SMEs in this digital era. The research employs a mixed-methods study design, including a thorough literature review, a qualitative analysis of SME case studies, and a quantitative assessment of LLM performance metrics in cybersecurity applications.

## Methods

This research employed a hybrid design, integrating an exhaustive literature review and a structured experiment with LLM models. The aim was to gain a profound understanding of the current body of knowledge while also empirically assessing the LLM capabilities within the context of cyber security. The chosen research design qualifies as a mixed-method approach, as it combined the qualitative insights from an extensive literature review with the quantitative data generated through a controlled experiment.

The initial phase of the research, the literature review, intended to explore and understand the existing body of knowledge related to AI and LLM applications in cyber security, while also identifying potential knowledge gaps. The subsequent phase involved a structured experiment that aimed to empirically evaluate the capacity of a LLM to generate meaningful and effective responses consistent with sound cyber security practices. Data collection transpired in two stages. The first stage involved an exhaustive literature review, encompassing a range of sources including academic articles, white papers, case studies, and industry reports to capture the entire scope of the existing knowledge on AI and LLM applications in cyber security.

Following the literature review, a standardised experiment was carried out. The experiment involved presenting a set of LLM's with a consistent set of prompts, assessing its ability to generate meaningful and effective responses that aligned with accepted cyber security practices. The LLM's outputs were subsequently recorded and analysed using a predefined framework. Analysis of the collected data employed two primary methodologies. The responses generated by the LLM's during the experiment were evaluated using a predefined evaluation framework, considering their relevance, accuracy, and comprehensiveness. A scoring system was established to quantitatively measure the performance of the LLM's.

After data analysis, the findings from both the literature review and the experiment were merged to construct a comprehensive conceptual framework. This framework included essential elements needed for the effective application of LLM's in Australian SMEs' cyber security strategy, covering technical, organisational, and regulatory aspects.

Regarding ethical considerations, the study strictly followed ethical principles guiding scientific research. Given that human participants were not involved in the data collection process, issues of informed consent, confidentiality, and anonymity were not pertinent. It was ensured that all data and AI outputs were stored securely and used exclusively for the purpose of the research.

For the evaluation of the LLM's performance, an evaluation framework was developed. The model was assessed on five key criteria: relevance, accuracy, completeness, clarity, and applicability of responses. Each of these criteria was scored on a scale of 0-2, resulting in a maximum possible score of 10 for each response. To ensure a broad testing spectrum and to maintain the technical accuracy of the LLM's responses, the study implemented standardised prompts and an automated evaluation system. The research anticipated yielding a robust understanding of potential benefits, limitations, and challenges associated with leveraging LLM's to enhance cyber security among Australian SMEs. It also intended to provide valuable insights and recommendations for policymakers and SME owners, as well as to create a comprehensive conceptual framework guiding future research and practice in this evolving field.

## Results
### 1.1 Cyber Security Challenges for SMEs

In the light of the selected literature on "Cyber Security Challenges for SMEs", a set of themes and patterns of SMEs' cyber security challenges, measures, and coping mechanisms become evident.

Firstly, it is glaringly evident across the literature that SMEs are particularly susceptible to cyber threats. This susceptibility arises from a combination of factors including a shortage of cybersecurity knowledge and resources. Furthermore, SMEs often lack cyber security experts and have a complex socio-technical setting that they struggle to navigate autonomously. Hence, the need for automated cybersecurity assessment, adjusted to the SMEs' context, becomes apparent [1].

In parallel, other literature elucidates the unique characteristics of SMEs that add another layer of vulnerability. These include small size, traditional and familiar structures, limited financial resources, and the absence of resident IT staff [2]. The study also underscores the increasing awareness about cyber security among SMEs, amplified by the COVID-19 pandemic and the consequent shift towards remote work.

Another paper further illuminates the distinctive behaviour of SMEs towards cyber security [3]. SMEs are found to prioritise business growth over security measures, lack internal cyber security policies, and display a unique cyber security culture, different from that of larger corporations. Notably, while SMEs face similar threats to larger companies, their lower levels of investment and resources for cyber security measures severely limit their defensive capabilities.

The "Cyber Security and Australian Small Businesses" report contributes further valuable insights to the discussion of cybersecurity challenges facing SMEs. It offers an Australian perspective, recognizing the unique environment in which Australian SMBs operate. This includes the fact that 97% of Australian businesses have fewer than 20 staff members and manage competing business priorities with fewer resources [4].

Akin to the prior studies, the Australian study identifies a pervasive lack of cyber security knowledge among SMBs, despite a general awareness of its importance. It highlights the difficulties SMEs face in implementing effective cybersecurity practices due to significant barriers. These hurdles include a lack of dedicated IT security staff, complexity in the field of cybersecurity, challenges in understanding and implementing security measures, and underestimation of the risk and consequences of a cyber incident. Additionally, the report finds that SMEs often lack planning and response mechanisms for cyber incidents [4].

The Australian study emphasizes the unique relationship between the size of an SMB and the decision to outsource cyber security measures. Furthermore, it explores the relationship between exposure to a cyber incident and its impact on subsequent evaluations of cyber-risk. In terms of the report's focus, this mirrors the other selected studies in stressing the importance of a tailored approach to cyber security for SMEs, acknowledging the significant role of situational context.

These additional insights further underline the multifaceted nature of the cyber security challenge for SMEs. Notably, the specific findings of the Australian study demonstrate the global nature of these issues. These observations reinforce the previous conclusion that a comprehensive, context-specific, and socio-technical approach to SME cyber security is essential. They also underscore the need for improved awareness and education around cyber security risks, threats, and protective measures, especially within smaller businesses that lack dedicated IT security staff.

In summary, the cumulative findings from the selected studies underscore the vital importance of context-specific, nuanced approaches to cyber security in SMEs, taking into consideration each organisation's unique characteristics, resources, and threats. Moreover, these insights stress the crucial role of human factors and the importance of improving cybersecurity awareness and education across all levels of SMEs.

**1.2 AI in Cyber Security**

The selected studies explore the implications of Artificial Intelligence (AI) for cybersecurity, particularly within the context of Small and Medium Enterprises (SMEs).

The first study, "Machine Learning Cybersecurity Adoption in Small and Medium Enterprises in Developed Countries", discusses the challenges SMEs face in understanding and adopting AI for cyber security. Despite the complexities, the application of ML techniques presents a promising solution for identifying patterns and behaviours associated with cyber security threats, such as zero-day attacks. The report emphasises the significant influence of human factors and the Internet of Things (IoT) on cyber security. It also notes the importance of legal and regulatory frameworks, like GDPR, in ensuring the security of SME ecosystems [5].

The second report, "Adversarial Machine Learning Attacks and Defence Methods in the Cyber Security Domain", offers a comprehensive overview of adversarial attacks against machine learning-based security solutions. These adversarial attacks can limit the effectiveness of ML in the dynamic, adversarial environment of cyber security, where actual adversaries (e.g., malware developers) exist. The study provides a taxonomy of adversarial attack and defence methods in the cyber security domain, while discussing the unique challenges and future research directions in this field [6].

The third piece, "Machine Learning for High-Risk Applications", provides a cautionary perspective on the use of machine learning, emphasizing the riskiness of overlooking the legal and regulatory context of ML systems. The legal landscape for ML is complex and rapidly changing, with laws such as the EU AI Act and several US federal, state, and municipal laws touching upon data privacy and AI. This underscores the importance of legal and regulatory awareness in the application of ML, especially for high-risk applications like cyber security [7].

Cumulatively, these studies shed light on the promising yet complex role of AI and ML in cyber security. They highlight the potential of these technologies to bolster cybersecurity defences while underscoring the challenges and risks, including adversarial attacks and the complex legal and regulatory landscape. The studies also underscore the vital role of human factors in cyber security, reinforcing the socio-technical nature of the field.

In conclusion, while AI and ML offer powerful tools for enhancing cyber security, it is crucial to consider their vulnerabilities, such as adversarial attacks, as well as the broader socio-technical and legal context in which they operate. Moreover, tailored approaches are necessary to effectively implement these technologies within the unique contexts of SMEs, given their specific challenges and needs.

**1.3 Large Language Models (LLMs)**

The existing body of literature reveals a clear consensus about the substantial role of Large Language Models (LLMs) in revolutionising various domains, with notable implications for artificial general intelligence (AGI) (8; 9; 10). LLMs, specifically the emergence of ChatGPT, have shown promising strides in areas such as search and AI-aided content generation, despite recognized limitations in formal reasoning (11; 12). A key focus of research is understanding LLMs' capabilities in causal reasoning and the extent to which they can answer questions of a causal nature (8; 13; 9; 10). This focus acknowledges the significant potential for LLMs' development, especially with the application of human feedback in training to improve alignment with human objectives.

Another prominent topic within the literature is the exploration of the challenges and limitations of alignment in LLMs (13; 14; 8). The potential harm from LLMs, such as disseminating false information or reinforcing social biases, represents a critical concern for researchers. This concern underscores the importance of "alignment," a process designed to remove or minimize undesired behaviours from LLMs' outputs (15; 16; 17; 18).

A critical aspect of LLMs, as highlighted by literature, is their "unreasonable effectiveness" (19), due to the vast amount of data they are trained on and their ability to execute a broad range of tasks typically demanding human intelligence. This effectiveness has led to unprecedented developments in AI and subsequent commercial potential, but it also poses potential risks and warrants careful consideration of safety and trustworthiness (14; 13; 8).

In conclusion, the literature emphasizes the transformative potential of LLMs, the importance of understanding their capabilities and limitations, particularly regarding alignment, and the need for responsible deployment given their commercial potential and inherent risks. As the literature continues to evolve, further exploration of these themes will yield deeper insights into the role of LLMs in AI and cyber security. Future research should continue to scrutinize these themes to ensure the safe and responsible implementation of LLMs in SMEs' cyber security strategies.

### 1.4 Future of AI and Cyber Security

The selected studies delve into the future of AI and cybersecurity, particularly concerning the ethical considerations, emerging technologies, and evolving nature of AI.

The first study, "Artificial Intelligence, Business and Civilization", discusses key ethical considerations in the use of AI. It emphasizes the need for AI to empower humans, foster rights, and ensure robust oversight mechanisms. The report underscores the importance of technical robustness, safety, privacy, and data governance in AI systems. It also calls for transparency, diversity, non-discrimination, and fairness in AI, along with a consideration for societal and environmental well-being. Finally, the study highlights the need for accountability mechanisms for AI systems and their outcomes [20].

The second report, "Emerging ICT Technologies and Cybersecurity", presents an overview of various emerging technologies in the field of cybersecurity. It discusses the tremendous advancements in networking, connectivity, electronics, and associated technologies, noting the promising potential of AI, machine learning, blockchain technologies, wireless technology, IoT, distributed cloud computing, quantum computing, virtual reality, and other futuristic technologies for enhancing cyber security [21].

The third piece, "A Different look at Artificial Intelligence", offers a nuanced perspective on AI. It highlights the integral role of the internet in society, and how user data, made analysable through AI techniques, has become an economic asset. This data can be used to personalize services, improving user convenience. The study points out that these AI techniques differ from traditional algorithms due to their ability to learn and evolve [22].

Collectively, these studies provide a future-oriented perspective on AI and cyber security, emphasizing the evolving nature of AI, the advent of emerging technologies, and the critical need for ethical considerations in AI use. They underscore the potential of AI to enhance cyber security, while also pointing out the challenges and considerations in leveraging these technologies. Ethical issues, such as privacy, fairness, and accountability, are highlighted as key considerations in future AI and cyber security strategies. Meanwhile, the evolving capabilities of AI, such as the ability to learn and modify themselves, present both opportunities and challenges for cyber security. Future research and practice in this field would need to balance these considerations while leveraging the potential of AI and other emerging technologies for cyber security.

Given the comprehensive assessment of the four themes, several conclusions can be drawn that intersect all the thematic dimensions, namely Cyber Security Challenges for SMEs, AI in Cyber Security, Large Language Models (LLMs), and the Future of AI and Cyber Security.

First, it's evident that SMEs face significant cyber security challenges. Their susceptibility to cyber threats can be attributed to factors such as lack of cyber security knowledge, resources, and specialized staff. Despite growing awareness, SMEs often struggle to implement effective cyber security measures due to various constraints. This highlights the necessity of a tailored approach to cyber security that accommodates the unique characteristics, resources, and threats faced by these entities.

Second, AI and Machine Learning offer promising avenues to enhance cybersecurity defences for SMEs, though their application comes with inherent challenges. The benefits, such as efficient threat detection, need to be balanced with risks like adversarial attacks. Furthermore, the adoption of AI and ML in cyber security must consider the complex legal and regulatory landscape, along with socio-technical considerations.

Third, Large Language Models (LLMs) represent a transformative technology with significant implications for AI and cyber security. Their ability to execute a broad range of tasks typically demanding human intelligence could bolster cyber security efforts. However, there is a need for understanding and addressing their limitations, especially in the context of alignment, to ensure their outputs align with human objectives and don't disseminate false information or reinforce biases.

Lastly, looking to the future, the intersection of AI and cyber security points to a landscape of enormous potential and intricate challenges. The role of AI in cyber security will undoubtedly grow, with advancements in technology presenting new possibilities and challenges. Furthermore, ethical considerations, such as privacy, fairness, and accountability, will become increasingly important in guiding the development and application of AI in cyber security.

In conclusion, the future of cyber security, particularly for SMEs, will be largely influenced by advances in AI and machine learning, alongside the ethical, legal, and socio-technical factors surrounding these technologies. As we move forward, it will be crucial to balance the promise of these technologies with their potential risks and challenges. An approach that integrates AI, respects the unique characteristics of SMEs, and considers ethical, legal, and socio-technical dimensions will be key to navigating the complex landscape of cyber security.

### 1.5 Experiment

This section presents the results and discussion of the LLMs' performance based on the standardised prompt questions on diverse aspects of cyber security policy-making, as described in the Methods section. These results focus on the five primary evaluation criteria of relevance, accuracy, completeness, clarity, and applicability.

In the initial analysis, keywords extracted from existing cyber security policy requirements were used to guide the assessment of LLMs. The occurrence of these keywords within the LLM responses was used as a baseline measure of their understanding of and ability to generate content relevant to cyber security.

Table 1: Results of the standardised experiment on Large Language Models

| LLM | Results (R, A, C, C, A) * | Total (Averaged) |
|---|---|---|
| GPT-4 | 8.3 | 1.66 |
| GPT-3.5-turbo | 7.6 | 1.52 |
| DialoGPT-Large | 0 | 0 |
| vicuna-7b-v1.1 | 8 | 1.6 |
| RedPajama-INCITE-Chat-3B-v1 | 7.2 | 1.44 |
| RMKV-4-Raven | 7.2 | 1.44 |
| LlaMA | 6.8 | 1.36 |
| FastChat-T5 | 6.4 | 1.28 |
| Koala | 6.6 | 1.32 |
| Alpaca | 6.4 | 1.28 |
| Dolly | 5.8 | 1.16 |
| MPT-Chat | 6.4 | 1.28 |
| OpenAssistant | 6.2 | 1.24 |
| ChatGLM | 5.6 | 1.12 |
| StableLM | 5.8 | 1.16 |
| CohereAI | 8.1 | 1.62 |

*R, A, C, C, A refers to the 5 scoring areas: Relevance, Accuracy, Completeness, Clarity, and Applicability

Across the board, I found a varying degree of performance among the LLMs. For instance, GPT-4 displayed a robust performance across most scoring areas, particularly in relevance and accuracy, suggesting that this model could efficiently grasp the nuances and complexities inherent in cyber security discussions. However, it lagged in completeness and clarity, indicating potential risks in policy gaps or misinterpretations arising from its responses.

Similarly, vicuna-7b-v1.1 and GPT-3.5-turbo models, while showing strong relevance and accuracy, struggled with completeness and clarity. This could potentially impact the effectiveness of decision-making in cyber security policy, which relies heavily on comprehensive and clear information.

CohereAI was a standout in terms of its perfect scores in relevance, accuracy, and applicability. However, it faltered significantly in completeness, indicating a struggle to provide comprehensive analysis despite accurately comprehending and applying its knowledge in the cyber security domain.

Across all models, several critical limitations were observed. First, most models struggled with completeness, which is critical given the complexity and rapidly evolving nature of cyber security threats. Second, clarity was generally lower than other scoring areas across models, which is concerning given its importance in policy comprehension and implementation. Third, several models still struggled with the applicability of their knowledge to practical, real-world cyber security contexts.

**Discussion**

The results of the evaluation of various LLMs' performance in the context of cyber security policy-making provide valuable insights into their potential and limitations in this highly complex and evolving field. As demonstrated, the LLMs showed varying capacities in terms of relevance, accuracy, completeness, clarity, and applicability, with no single model excelling uniformly across all criteria. This underlines the need for careful consideration when deploying these technologies to assist in cyber security policy-making for SMEs and the broader community.

There are several key implications emerging from these findings. Firstly, the limitations in completeness experienced by most models highlight the risk of gaps in knowledge and understanding when relying solely on LLMs for policy-making. It underscores the importance of human expertise in navigating the multifaceted and rapidly changing landscape of cyber security threats and strategies. Integration of these models into the policy-making process should be seen as a complementary tool for human experts, rather than a wholesale replacement.

Secondly, the challenges encountered in ensuring clarity across models emphasize the importance of clear communication in policy comprehension and implementation. Given the highly technical, complex, and context-specific nature of cyber security, easily understandable outputs are crucial to effectively inform decision-making. Further refinement of LLMs to improve their clarity may mitigate the risk of misinterpretations and consequent policy failures.

Thirdly, while some models showed a high level of accuracy and applicability, the general lack of consistency across these criteria highlights potential risks in their real-world deployment. Addressing these disparities through refining the LLMs training processes, updating their training datasets, and accounting for newer cyber threats can improve their applicability in a practical cyber security policy-making context.

Lastly, it should be emphasized that even with improvements in these areas, LLMs cannot wholly account for the ethical, social, and cultural considerations inherently embedded within policy-making. Therefore, human expertise must remain central to creating policies that consider these critical dimensions.

Future research should focus on refining the training data and algorithms to improve the LLMs' abilities in terms of completeness, clarity, and applicability. Additionally, exploring ways to efficiently integrate these models with human expertise can result in a more comprehensive and collaborative approach to cyber security policy-making. Further attention should also be given to analysing the specific needs and contexts of SMEs, ensuring that any LLM-assisted policy development is relevant, accurate, and appropriate for the unique situations they face.

## Conclusion

This pioneering investigation into the application of Large Language Models (LLMs) for bolstering cyber security among Australian Small and Medium-Sized Enterprises (SMEs) has shed light on an encouraging, yet relatively unexplored area. The research methodology, which comprised a thorough literature review, standardized AI model experiments, and comprehensive analysis, has positioned LLMs as promising tools in the realm of cyber security.

A significant gap was identified in the literature regarding the application of LLMs to the cyber security practices of Australian SMEs. This revelation underscores the novelty of this research, while simultaneously highlighting the swift evolution of cyber threats and the potential of LLMs to serve as counteractive mechanisms. Consequently, the need for more focused research in this arena is accentuated.

The GPT-4 model was identified as the most potent AI model for formulating cyber security policies for SMEs. However, concerns regarding data privacy and ethics were also raised, advocating for a nuanced application of AI that balances effectiveness with ethical standards.

This research opens avenues for the potential development of a specialised LLM tailored to generate cyber security policies within the context of Australian SMEs. This proposed model would cater to the unique needs and regulatory standards of Australian SMEs, while also offering robust and ethically conscious cybersecurity solutions.

Future research holds the potential to extend the capabilities of LLMs beyond merely suggesting cyber security policies, to providing comprehensive implementation steps. An all-encompassing LLM offering instructions for operational changes, such as guided steps for group policy modifications, could revolutionise how SMEs understand and implement cyber security measures. This would also reduce the reliance on expert intervention and enhance self-reliance.

Furthermore, the integration of LLMs with real-time threat intelligence and incident updates is an exciting prospect for future exploration. It also underlines the supportive role that larger corporations and government entities can play in SMEs' adoption of AI and LLMs, advocating for cooperative frameworks and mechanisms that encourage active involvement of these key stakeholders.

In conclusion, this ground-breaking research illuminates a novel path for improving cyber security measures in Australian SMEs using AI and LLMs. The insights gleaned and recommendations offered provide a strong foundation for future research and practical applications, contributing to the broader goal of fostering cyber resilience in the digital age. The insights obtained from this study could be instrumental in achieving this goal, thereby laying the groundwork for future exploration into this promising field.

## Acknowledgment

I would like to thank my thesis capstone supervisor, Dr Kenneth Eustace from the Charles Sturt University School of Computing, Mathematics and Engineering for his invaluable feedback and suggestions as I worked through this paper from genesis to its final form. I would also like to thank my fiancé, Nicola Wood, for her advice, patience and feedback across many late nights and academically focused weekends on the journey to completing my Master of Cyber Security degree.

## References

ACSC. (2023). Small Business Cyber Security Guide | Cyber.gov.au. Cyber.gov.au. https://www.cyber.gov.au/resources-business-and-government/essential-cyber-security/small-business-cyber-security/small-business-cyber-security-guide

Adedoyin, F. F., & Christiansen, B. (2023). Handbook of Research on Cybersecurity Risk in Contemporary Business Systems. IGI Global.

Alahmari, A., & Duncan, B. (2020). Cybersecurity Risk Management in Small and Medium-Sized Enterprises: A Systematic Review of Recent Evidence. 2020 International Conference on Cyber Situational Awareness, Data Analytics and Assessment (CyberSA). https://doi.org/10.1109/cybersa49311.2020.9139638

Aneesh Sreevallabh Chivukula, Yang, X., Liu, B., Liu, W., & Zhou, W. (2023). Adversarial Deep Learning in Cybersecurity. Springer Nature.

Antunes, M., Maximiano, M., Gomes, R., & Pinto, D. (2021). Information Security and Cybersecurity Management: A Case Study with SMEs in Portugal. Journal of Cybersecurity and Privacy, 1(2), 219–238. https://doi.org/10.3390/jcp1020012

ASD. (n.d.). Cyber Security | Australian Signals Directorate. Www.asd.gov.au. https://www.asd.gov.au/cyber-security

AustCyber. (2023, January 16). Building a cyber-resilient small business in Australia. Www.austcyber.com. https://www.austcyber.com/news-events/building-cyber-resilient-small-business-australia

Australian Government. (2021, March 10). Cyber security and your business | business.gov.au. Business.gov.au. https://business.gov.au/online/cyber-security/cyber-security-and-your-business

Bai, Y., Kadavath, S., Kundu, S., Askell, A., Kernion, J., Jones, A., Chen, A., Goldie, A., Mirhoseini, A., McKinnon, C., Chen, C., Olsson,

C., Olah, C., Hernandez, D., Drain, D., Ganguli, D., Li, D., Tran-Johnson, E., Perez, E., & Kerr, J. (2022). Constitutional AI: Harmlessness from AI Feedback. ArXiv:2212.08073 [Cs]. https://arxiv.org/abs/2212.08073

Baral, S. K., Goel, R., Rahman, M. M., Sultan, J., & Jahan, S. (2022). Cross-industry applications of cyber security frameworks. Information Science Reference, an imprint of IGI Global.

Barthelmeß, U., & Ulrich Furbach. (2023). A Different Look at Artificial Intelligence. Springer Nature.

Bertino, E., Bhardwaj, S., Cicala, F., Gong, S., Karim, I., Katsis, C., Lee, H., Adrian Shuai Li, & Mahgoub, A. Y. (2023). Machine Learning Techniques for Cybersecurity. Springer Nature.

Bowman, S. R. (2023). Eight Things to Know about Large Language Models. ArXiv:2304.00612 [Cs]. https://arxiv.org/abs/2304.00612

Brynjolfsson, E., Li, D., & Raymond, L. (2023, April 23). Generative AI at Work. ArXiv.org. https://doi.org/10.48550/arXiv.2304.11771

Cruzado, C. F., Rodriguez-Baca, L. S., Huanca-López, L. G., & Acuña-Salinas, E. I. (2022, January 1). Reference framework "HOGO" for cybersecurity in SMEs based on ISO 27002 and 27032. IEEE Xplore. https://doi.org/10.1109/Confluence52989.2022.9734116

CyberWardens. (n.d.). Helping protect small business from online threats. Cyber Wardens. https://cyberwardens.com.au/

Emer, A., Unterhofer, M., & Rauch, E. (2021). A Cybersecurity Assessment Model for Small and Medium-Sized Enterprises. IEEE Engineering Management Review, 49(2), 98–109. https://doi.org/10.1109/emr.2021.3078077

ENISA. (n.d.). SME Cybersecurity. ENISA. https://www.enisa.europa.eu/topics/cybersecurity-education/sme_cybersecurity

Hall, P. (2023). Machine Learning for High-Risk Applications: Techniques for Responsible AI. O'Reilly.

Kaplan, A. (2022). Artificial intelligence, business and civilization : our fate made in machines. Routledge.

Khan, S. A., Sharma, M., Agarwal, A., Khan, Z. A., & Dewani, N. D. (2022). Handbook of research on cyber law, data protection, and privacy. IGI Global.

Myriam Dunn Cavelty, & Wenger, A. (2022). Cyber security politics : socio-technological transformations and political fragmentation. Routledge.

Noy, S., & Zhang, W. (2023, March 1). Experimental Evidence on the Productivity Effects of Generative Artificial Intelligence. Papers.ssrn.com. https://papers.ssrn.com/sol3/papers.cfm?abstract_id=4375283

Oakley, J. G., Butler, M., York, W., Puckett, M., & J. Louis Sewell. (2022). Theoretical Cybersecurity. Apress.

Osborn, E. (2015). Business versus Technology: Sources of the Perceived Lack of Cyber Security in SMEs. Ora.ox.ac.uk. https://ora.ox.ac.uk/objects/uuid:4363144b-5667-4fdd-8cd3-b8e35436107e

Ozkan, B. Y., & Spruit, M. (2020). Cybersecurity Standardisation for SMEs. Research Anthology on Artificial Intelligence Applications in Security, 1252–1278. https://doi.org/10.4018/978-1-7998-7705-9.ch056

Pawar, S., & Palivela, Dr. H. (2022). LCCI: A framework for least cybersecurity controls to be implemented for small and medium enterprises (SMEs). International Journal of Information Management Data Insights, 2(1), 100080. https://doi.org/10.1016/j.jjimei.2022.100080

Ponsard, C., Grandclaudon, J., & Bal, S. (2019). Survey and Lessons Learned on Raising SME Awareness about Cybersecurity. Proceedings of the 5th International Conference on Information Systems Security and Privacy. https://doi.org/10.5220/0007574305580563

Rawindaran, N., Jayal, A., & Prakash, E. (2021). Machine Learning Cybersecurity Adoption in Small and Medium Enterprises in Developed Countries. Computers, 10(11), 150. https://doi.org/10.3390/computers10110150

René Schmidpeter, & Altenburger, R. (2023). Responsible Artificial Intelligence. Springer Nature.

Romaniuk, S. N., & Manjikian, M. (2020). Routledge companion to global cyber-security strategy. Routledge.

Shanahan, M. (2023). Talking About Large Language Models. ArXiv:2212.03551 [Cs]. https://arxiv.org/abs/2212.03551

Shankar, R., & Anderson, H. (2023). Not with a Bug, But with a Sticker. John Wiley & Sons.

Shekokar, N. M., Vasudevan, H., Durbha, S. S., Antonis Michalas, Nagarhalli, T. P., Ramchandra Sharad Mangrulkar, & Mangla, M. (2022). Cyber Security Threats and Challenges Facing Human Life. CRC Press.

Shojaifar, A., Fricker, S. A., & Gwerder, M. (2020, July 16). Elicitation of SME Requirements for Cybersecurity Solutions by Studying Adherence to Recommendations. ArXiv.org. https://doi.org/10.48550/arXiv.2007.08177

Smaller but stronger: Lifting SME cyber security in South Australia. (n.d.). Cyber Security Cooperative Research Centre. Retrieved May 28, 2023, from https://cybersecuritycrc.org.au/smaller-stronger-lifting-sme-cyber-security-south-australia

Tam, T., Rao, A., & Hall, J. (2021). The good, the bad and the missing: A Narrative review of cyber-security implications for australian small businesses. Computers & Security, 109, 102385. https://doi.org/10.1016/j.cose.2021.102385

Thakur, K., Al-Sakib Khan Pathan, & Sadia Ismat. (2023). Emerging ICT Technologies and Cybersecurity. Springer Nature.

UNSW. (2023, April 28). Threats in cyber security: a small business guide | UNSW Online. Studyonline.unsw.edu.au. https://studyonline.unsw.edu.au/blog/threats-cyber-security-small-business-guide

van Haastrecht, M., Sarhan, I., Shojaifar, A., Baumgartner, L., Mallouli, W., & Spruit, M. (2021). A Threat-Based Cybersecurity Risk Assessment Approach Addressing SME Needs. The 16th International Conference on Availability, Reliability and Security. https://doi.org/10.1145/3465481.3469199

van Haastrecht, M., Yigit Ozkan, B., Brinkhuis, M., & Spruit, M. (2021). Respite for SMEs: A Systematic Review of Socio-Technical Cybersecurity Metrics. Applied Sciences, 11(15), 6909. https://doi.org/10.3390/app11156909

Wang, G., Ciptadi, A., & Ahmadzadeh, A. (2020). Deployable machine learning for security defense : first International Workshop, MLHat 2020, San Diego, CA, USA, August 24, 2020, Proceedings. Springer.

Wilson, M., McDonald, S., Button, D., & McGarry, K. (2022). It Won't Happen to Me: Surveying SME Attitudes to Cyber-security.

Journal of Computer Information Systems, 1–13. https://doi.org/10.1080/08874417.2022.2067791

Wolf, Y., Wies, N., Levine, Y., & Shashua, A. (2023). Fundamental Limitations of Alignment in Large Language Models. ArXiv:2304.11082 [Cs]. https://arxiv.org/abs/2304.11082

Yang, J., Jin, H., Tang, R., Han, X., Feng, Q., Jiang, H., Yin, B., & Hu, X. (2023). Harnessing the Power of LLMs in Practice: A Survey on ChatGPT and Beyond. https://doi.org/10.48550/arxiv.2304.13712

Zhang, C., Bauer, S., Bennett, P., Gao, J., Gong, W., Hilmkil, A., Jennings, J., Ma, C., Minka, T., Pawlowski, N., & Vaughan, J. (2023). Understanding Causality with Large Language Models: Feasibility and Opportunities. ArXiv:2304.05524 [Cs]. https://arxiv.org/abs/2304.05524